\documentclass[a4paper,fleqn,usenatbib]{mnras}

\usepackage{newtxtext,newtxmath}
\newcommand{\kms}{\,km\,s$^{-1}$}
\newcommand{\ms}{\,m\,s$^{-1}$ }

\usepackage[T1]{fontenc}
\usepackage{ae,aecompl}

\usepackage{enumitem}
\usepackage[caption=false]{subfig}
\usepackage{amssymb}
\usepackage{amsmath}
\usepackage{commath}
\usepackage{graphicx,bm}
\usepackage{verbatim}

\usepackage{graphicx}	
\usepackage{amsmath}	
\usepackage{amssymb}	
\usepackage[utf8]{inputenc}

\title[Detection of planetary signals by reflected light of the host star using the ACF]{Detection of planetary signals by reflected light of the host star using the autocorrelation of spectra }

\author[E.F. Borra \& D. Deschatelets]{
E.F. Borra \& D. Deschatelets \thanks{E-mail: \url{borra@phy.ulaval.ca}, \url{david.deschatelets.1@ulaval.ca}}
\\
Département de Physique, Université Laval, Québec, Qc, Canada G1V 0A6 
}

\date{Accepted XXX. Received YYY; in original form ZZZ}

\pubyear{2016}

\begin{document}
\label{firstpage}
\pagerange{\pageref{firstpage}--\pageref{lastpage}}
\maketitle

\begin{abstract}

We consider an alternative to the cross-correlation function (CCF), that uses the autocorrelation function (ACF), to measure in spectra the reflected light of the stars by their planetary companion. The objective of this work is to assess and analyse the efficiency of the ACF in detecting planetary signals by a detection of reflected light. To do so, we first compare the ACF and the CCF using artificial spectra containing a planetary signal. We then use the ACF to analyse spectra of the 51 Peg + 51 Peg b system and compare our results with those obtained by \cite{martins2015} who previously analysed the same system using the CCF. The functionalities of the ACF and the way it is implemented are similar to that of the CCF. One of the main advantages of the ACF is the fact that, unlike the CCF, no weighted binary mask is required. This makes the ACF easier to use. The results related to simulated spectra showed that the ACF allowed us to decrease the boundary condition threshold for the use of spectra compared to the CCF so that more data could be used to recover the planetary signal. For the 51 Peg b planet, we achieved a detection significance of 5.52 $\sigma_\textrm{noise}$ with the ACF compared to 3.70 $\sigma_\textrm{noise}$ with the CCF \citep{martins2015}. We conclude that the ACF has the potential to become a prominent technique in detecting exoplanets considering its efficiency, ease of use and rapid execution time.    

\end{abstract}




\section{Introduction} 

Until recently, planetary signals were detected by means of methods such as transit photometry and radial velocity \citep[e.g.][]{charbonneau2000,queloz2000,reiners2010}. The theory underlying the subject of exoplanets has been covered extensively in previous works \citep[e.g.][]{charbonneau1999,marley1999,stam2004}. Attempts in detecting reflected light were first made by analysing giant planets orbiting near their host star. For instance, \cite{charbonneau1999} attempted to no avail to find evidence of reflected light related to the planet orbiting the host star $\tau$ Boo using the HIRES echelle spectrograph. Only upper limits pertaining to the planet-to-star flux ratio as well as the planetary geometric albedo in the optical band were determined. More recently, \cite{rodler2010} undertook a similar work by searching for reflected light associated with the presence of $\tau$ Boo b using high-resolution spectra collected with the UVES optical spectrograph. Although the planetary signal was too weak to be declared a detection, they were able to establish precise upper limits to the planet-to-star flux ratio.

 \cite{martins2015} used a technique involving the cross-correlation function (CCF) of spectra to detect the planetary signal of 51 Peg b using visible light reflection.  This method is discussed at length in \cite{martins2013}. Measuring the signal of the planet directly in the visible spectrum is challenging due to the noise and the weak planet-to-star flux ratio of around $10^{-5}$ to $10^{-4}$ \citep[e.g.][]{seager2000,sudarsky2003,rowe2008}. The 51 Peg b planet discovered by \cite{mayor1995} has been studied in previous works \citep[e.g.][]{brogi2013,martins2015,birkby2017}. In the optical band, the planetary signal is mainly characterized by reflected light of its host star. Thus, the spectrum of a planet in this band is a copy of the stellar signal but with a much weaker intensity. In the near-infrared, the signal of the planet is mostly dominated by thermal emission which can be problematic for the analysis \citep{marley1999}. Moreover, a greater number of spectral lines can be found in the visible band of the spectra for the analysis. Detecting a planetary signal directly with reflected light allows one to measure its orbital speed, from which other parameters such as the mass of the planet can be determined.

The objective of this work is to assess the possibility to apply the autocorrelation function (ACF) to spectra of stars + planets systems and recover the spectroscopic signal of the planet by means of a direct detection of reflected light. We analyse the efficiency of the ACF in recovering weak planetary signals and compare it to the CCF which is an already well-established method in the domain. This paper primarily focuses on comparing results obtained with the ACF and the CCF by using simulated spectra and data to analyse the 51 Peg + 51 Peg b planetary system.     
 
 The ACF shares many similarities with the CCF and therefore the way both methods are implemented to detect planetary signals is comparable. The autocorrelation function was first introduced in our previous works to measure the variation in the amplitude of magnetic fields as a function of the rotation phase of magnetic stars (\cite{borradeschatelets2015}). It was also used in a subsequent work to measure and quantify microturbulence motion as a function of the pulsational phase of Cepheids (\cite{borradeschatelets2017}). In each of these astrophysical fields, the ACF was compared with other techniques and provided us with better results. The ACF has some advantages over the CCF that facilitate the process of detecting planetary signals. Therefore, it is a tool that lends itself well to the study of exoplanets and the determination of their physical characteristics. 

In Section \ref{reflectedlight}, we review some basic theory elements related to exoplanets and their orbital properties. We discuss the properties of the ACF in Section \ref{autocorr} and the way it is implemented in our work for the analysis of exoplanets. We also lay out the advantages of using the ACF to detect planetary signals compared to the CCF in that section. In Section \ref{method}, we present the methodology behind the ACF using simulated spectra. The same protocol is followed when analysing real data. In Section \ref{results}, we first present the results and compare them between the ACF and the CCF using simulated spectra containing a planetary signal. We then present the results obtained with the ACF for the 51 Peg + 51 Peg b system and compare the recovered planetary signal to that of \cite{martins2015} who used the CCF. We finally conclude the paper in Section \ref{conclusion}.

\section{Reflected light of planets}
\label{reflectedlight}

The spectrum of a planet is theoretically a copy of that of its host star. However, the reflected light of the planet produces a flux that is a small fraction of the intensity of the stellar signal. 

The orbital phase $\phi$ allows us to trace the positioning of the planet around the star. This parameter takes values between 0 and 1. The position of the orbital phase $\phi$ = 0 is not universal in astrophysics. Some authors define $\phi$ = 0 as the phase where the measured radial velocity of the planet is maximal while others define this phase as the moment when the planet is in transit in front of the star relative to the observer. In this work, we use the latter definition. At each instant $t$, we can calculate the orbital phase of the planet according to $\phi = \frac{t-t_0}{P_\textrm{orb}}$, where $t_0$ is the time of transit of the planet and $P_\textrm{orb}$ is its orbital period \citep{martins2013}. The portion of the hemisphere of the planet that is illuminated with respect to the observer varies according to the phase. The phase angle, represented by $\alpha$, characterizes the star-planet-observer angle. The phase angle takes values between $0 \leq \alpha \leq 180$° and depends on the position of the planet in its orbit. It can be calculated relatively to the orbital period with the following equation \citep{martins2013}:

\begin{equation}
cos(\alpha) = sin(I)cos(2\pi \phi).
\end{equation}

\noindent where $I$ is the orbital inclination angle with respect to the observer.

The phase function varies with the phase angle $\alpha$, which is the angle between the star and the Earth when seen from the planet. If we assume that the light reflection of the planet follows a Lambert sphere, we can define the phase function as follows \citep{cahoy2010}:

\begin{equation}
\Phi(\alpha) = \frac{sin(\alpha)+(\pi - \alpha)cos(\alpha)}{\pi}.
\label{1.8}
\end{equation}

The Lambert sphere is a theoretical model that assumes an isotropic scattering of the light received over the entire portion of the visible hemisphere of the planet. When the planet is at the phase of superior conjunction, the phase angle $\alpha = 0$ and the phase function $\Phi(\alpha) $ reaches its maximum normalized value of 1.  We can finally obtain the ratio of fluxes between the planet and the star \citep{madhusudhan2012}:

\begin{equation}
\frac{F_{\textrm{p}}(\alpha)}{F_{\star}} =  \bigg(\frac{R_{\textrm{p}}}{a}\bigg)^2 A_g \Phi(\alpha),
\end{equation}

\noindent where $R_{\textrm{p}}$ is the radius of the planet, a is the orbital distance of the planet and $A_g$ is the geometric albedo of the planet.

\section{Autocorrelation of the spectrum}
\label{autocorr}

The autocorrelation function is a powerful and easy to use tool with properties similar to those of the cross-correlation function. As opposed to the CCF, the ACF does not require the use of a weighted binary mask which is a theoretical spectrum characterized by slits which aim to match the lines of a stellar spectrum as best as possible. The weighting of the mask slits depends on the intensity of the lines of the spectrum at the corresponding positions. This is an important advantage that makes the ACF easier to use.

The CCF with a weighted binary mask has proved to be a very efficient method in several fields of astronomy \citep[e.g.][in polarization spectroscopy and radial velocity measurements respectively]{kochukhov2010,hartmann2015}. For most spectrographs, only a few masks are available for a limited amount of spectral types (G2 being the most commonly used spectral type). Using a G2 mask for a star of a different subtype may result in a situation where the mask slits are not properly weighted with respect to the spectral lines. If a mismatch occurs between the weighted binary mask and the spectrum, some very weak lines could be overweighted which would cause noise amplification in the CCF profile. On the other hand, by using the autocorrelation of the spectra, we ensure a perfect match between the signal and the mask since in the ACF the mask is the signal itself. Therefore, with the ACF, we eliminate several potential sources of contamination that may arise under the use of the cross-correlation technique. These sources of contamination include improperly weighted lines in the CCF profile and missing lines in the mask. Furthermore, some slits in the mask may be positioned where there are no lines in the spectra but only continuum noise. These arguments become more prominent as the subtype of the stellar spectrum differs from the mask used to create the CCF profile.

The theory behind the ACF has already been covered in \cite{borradeschatelets2015} and \cite{borradeschatelets2017}. We summarize some important elements of theory behind the ACF in what follows. The autocorrelation of the intensity $I(\textrm{v})$ as a function of the velocity $\textrm{v}$ of the spectra is given by
\begin{equation}
I \otimes I = \int_{-\infty}^\infty I(\textrm{v} + \textrm{v'})I(\textrm{v})d\textrm{v}.
\label{eq1}
\end{equation}

This allows us to calculate an average line profile with a strong signal-to-noise ratio (SNR). The application of the ACF to the spectra is straightforward. Only a few minor changes need to be made to the spectra. Prior to using the ACF, it is necessary to remove some problematic regions in the spectra that are likely to render the ACF inaccurate because they contain strong lines (e.g. H lines, telluric lines). This can easily be done by setting the intensity to 0.0 in  wavelength intervals where such lines are present. The contribution of these lines to the autocorrelation is thereby totally eliminated. The idea behind this process is to obtain a line profile that is representative of all the lines contained in the spectrum of the star.

The autocorrelation of a Gaussian signal remains Gaussian with a FWHM increased by a factor $\sqrt{2}$. If we approximate the spectral lines to be perfect Gaussian functions, the ACF would increase the FWHM of the signal of the star and that of the planet by a factor of $\sqrt{2}$ when compared to the measurements obtained with the CCF technique. Given the actual shapes of spectral lines, this approximation is very good and the multiplicative factor of $\sqrt{2}$ can be taken in consideration when comparing results obtained between the ACF and the CCF.

\section{Methodology}
\label{method}

The method presented in this section to detect the signal of an exoplanet resembles in many ways the one introduced by \cite{martins2013} that uses the CCF. The CCF allowed them to recover the planetary signal of 51 Peg b with a significant level of detection \citep{martins2015}. For a more in-depth look at the methodology and the theory underlying star + planet systems for the CCF technique, we advise the reader to seek further information from \cite{martins2013} and \cite{martins2015}. Despite the similarity between the ACF and the CCF, there are multiple differences that are worth mentioning. We lay out the important points of this in what follows.

\subsection{Generating simulated spectra}
\label{generate}

For the simulations of stellar spectra, we use an artificial model that we used for previous works \citep{borradeschatelets2015,borradeschatelets2017}. The spectra are simulated with MATLAB software and contain 600 lines. The spectra are built with fewer lines and a better SNR in order to minimize the execution time because a considerable amount of execution time is required. This is the reason why we decided to use this particular artificial model for our data analysis. The spectral lines of each spectrum are represented by Gaussian functions with a FWHM of 7.50 \kms and a normalized intensity varying between -1.0 and 0.0 (the continuum is positioned at intensity 0.0 after being removed). In total, 30 spectra are simulated with an identical positioning of stellar lines. 20 of these spectra are cases where no planet signal is added. This corresponds to the phase in which the planet is in transit in front of the star (i.e. the phase when the planet lies between the star and the observer). At this phase, no reflected light from the planet is emitted to the observer. The results obtained with the ACF are affected by the number of lines involved in the calculation. The SNR of the signal obtained with the ACF is proportional to $\sqrt{N}$, where N is the number of lines used in the analysis. Considering the fact that this is identical to the way the CCF operates, the number of lines used in the simulated spectra does not benefit one technique over another.

The intensity of the added planetary signal is $5 \times 10^{-5}$ times that of the signal of the star which is around the value found for the 51 Peg + 51 Peg b system. The simulated planet has an orbital velocity semi-amplitude $K_{\textrm{planet}}$ of 125 \kms. The $\frac{M_\textrm{star}}{M_\textrm{planet}}$ ratio is set to 2500 which gives a $K_{\textrm{star}}$ of 50 \ms.

The spectra are first created in intensity $I(\lambda)$ as a function of wavelength units (Å) in order to simulate spectra commonly found in databases. We set the resolution of the simulated spectra to 0.01 Å per pixel which is the same as the HARPS spectrograph that has a spectral resolution of R $\sim$ 115 000 \citep{harps}. The following steps of the methodology are used for the simulated spectra as well as real spectra that were used for the analysis of the 51 Peg + 51 Peg b system in Section \ref{51peg}.

The spectra are originally defined with a fixed wavelength increment of 0.01 Å per pixel. We need to convert the spectra in velocity $I(\textrm{v})$ units because we want to use the autocorrelation to measure the observed radial velocity of the planetary signal in \kms. To do so, we interpolate the intensity value of each pixel over a new domain of definition in which there is now a fixed increment value in velocity (instead of a fixed increment value in wavelength) for each subsequent pixel. This new domain of definition is generated according to Eq. \eqref{eq1a}:
\begin{equation}
\lambda(i+1) = \lambda(i) + \lambda(i)\frac{\Delta \textrm{v}}{c}
\label{eq1a}
\end{equation}

\noindent where $\lambda(i)$ is the wavelength at the $i^{\textrm{th}}$ pixel, $\Delta \textrm{v}$ is the fixed velocity increment per pixel and $c$ is the speed of light.

The starting value at position $\lambda(1)$ is 3781.91 Å (which is around the value of the  HARPS spectrograph). The precision of the interpolation for consecutive pixels depends on the value of the velocity increment $\Delta \textrm{v}$. The computation time for this interpolation is inversely proportional to the value of the velocity increment set by the user. It is therefore favourable to use a value that optimize both precision and speed of calculations. We analyse the impact of the velocity increment value on both the ACF and the CCF in Section \ref{results}.

\subsection{Analysis of the CCF and ACF}
\label{whyuseacf}

The radial velocity of the signal of the star varies with the phase $\phi$. This relation is given by the following equation \citep{martins2013}:

\begin{equation}
 RV_\textrm{star} = -K_{\textrm{star}} sin(2\pi\phi)
 \end{equation} 
 
\noindent where $\textrm{RV}_\textrm{star}$ is the observed radial velocity of the star, $K_{\textrm{star}}$ is the orbital velocity semi-amplitude of the star and $\phi$ is the orbital phase of the star.

The amplitude of this variation depends on the characteristics of the orbiting planet. The necessary steps to directly detect the light reflected by a planet with the CCF technique can be roughly summarized as follows \citep{martins2015}: 

(i) First obtain CCFs of each spectrum by cross-correlating the spectra with a weighted binary mask of the appropriate spectral type,

(ii) Shift each CCF by the radial velocity of the star to center the signal of the star to the position of 0 \kms,

(iii) Build a high SNR template by stacking the CCFs obtained after having been corrected for their corresponding stellar radial velocity,

(iv) Divide one after the other the CCFs containing a planetary signal by the template built at step (iii) to remove the signal of the star,

(v) Shift each of the residual by the radial velocity of the planet to center each planetary signal to the position of 0 \kms,

(vi) Optimize the significance of the detection of the planet by stacking all of the residuals obtained at step (v).

The precision of the different shifts depends on the value of the velocity increment $\Delta \textrm{v}$ discussed in the previous section. It is important to use a value sufficiently low to ensure each CCF will properly stack when building the template (step iii). Otherwise, a significant amount of artifacts can be generated during step (iv) if the CCFs used to build the template are not perfectly aligned. For that reason, the efficiency of the CCF in measuring a planetary signal can be restricted by the precision of the interpolation (i.e. the velocity increment value $\Delta \textrm{v}$ set by the user).

On the other hand, when using the autocorrelation, the position of the signal of the star is always perfectly centered at 0 \kms because we correlate the spectra with themselves. Thus, as opposed the CCF technique, no shifting is required to take into account the radial velocity of the star with the ACF. We can therefore skip step (ii) entirely. In doing so, we totally eliminate the risk of generating shift-induced artifacts during step (iv). The ACF is thereby less sensitive to the precision of the interpolation which makes it particularly well suited for the search of exoplanets by means of a direct detection of reflected light.

\begin{figure}

  \includegraphics[width=\linewidth]{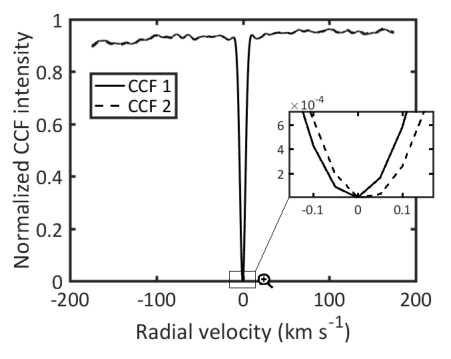}%

\caption{Velocity offset between two CCFs of the star 51 Peg for an interpolation with a velocity increment of 50 \ms per pixel.}
\label{fig2}
\end{figure}

CCFs of stellar spectra may have a very small offset of less than one pixel with respect to each others at the center value of 0 after having been corrected for the radial velocity of the system if the velocity increment per pixel used for the spectral analysis is set too high. When using the CCF to detect a planetary signal in spectra, this offset can become problematic. The velocity offset can be of more than 20 \ms depending on the velocity increment per pixel used and on the two spectra that are compared. \figref{fig2} shows the observed velocity offset between two normalized CCF profiles for the star + planet system 51 Peg + 51 Peg b. We can observe an offset of nearly 25 \ms between the two CCFs. The CCFs for 51 Peg were obtained by \cite{martins2015} who used the HARPS spectrograph to acquire the spectra. The CCFs that they used were interpolated to achieve a resolution of 50 \ms per pixel. Thereby, the observed velocity offset between the two CCFs on \figref{fig2} is around half a pixel.   

\begin{figure}

  \includegraphics[width=\linewidth]{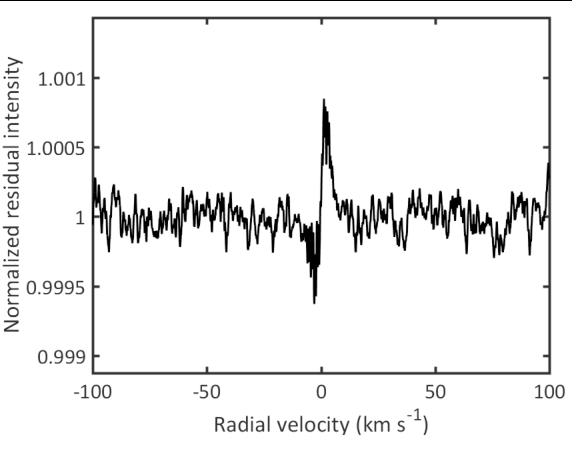}%

\caption{Residual intensity resulting from the division of a star + planet CCF by the star-only CCFs template for 51 Peg.}
\label{fig3}
\end{figure}

Small offsets between star + planet CCFs and the star-only CCFs template can lead, after division, to the introduction of strong artifacts at the position of the signal of the star. This is shown in \figref{fig3} for the case of 51 Peg. Stronger noise at position 0 is the consequence of the small mismatch between the stellar signal of the star + planet CCF and that of the star-only CCFs template. A solution to this problem, when using the CCF technique, is to avoid using CCFs at phases where the planetary signal is close to the centered signal of the star. Otherwise, the planetary signal would be suppressed by the presence of strong residual noise at position 0. For the 51 Peg + 51 Peg b system, \cite{martins2015} only used star + planet CCFs of spectra whose planetary signal is positioned at a minimal distance of $8 \times \textrm{FWHM}_{\textrm{star}}$ from the signal of the star. Although conservative, this condition allowed them to ensure that the planetary signal was not polluted by the strong residual noise originating from the small discrepancy between the signal of the star and the star-only CCFs template.

As mentioned earlier in this section, the autocorrelation of a stellar spectrum is the correlation of the spectrum with itself. As a result, the ACF profile is always symmetrical and well centered at position 0, regardless of the phase-dependant radial velocity of the system and the precision of the interpolation. Therefore, the use of the ACF allows us to limit the appearance of additional noise at the central position when dividing a star + planet ACF profile by the star-only ACFs template. The ACF suitably gives us the opportunity to use spectra at phases where the planet is near the superior conjunction. This provides the ACF a great advantage over the CCF method because we do not have to limit ourselves to spectra whose planetary signal is positioned at a minimum distance threshold with respect to the signal of the star. Also, the amount of light from the star that is reflected by the planet is maximal near the superior conjunction. This gives a brighter planetary spectrum with a greater signal-to-noise ratio. Therefore, more star + planet spectra can be used with the ACF in our analysis which allows us to take into account a larger amount of reflected light and improve the SNR of the measured planetary signal.

\subsection{Detecting the planetary signal with the ACF}
\label{detecting}

\begin{figure*}
\subfloat[Star + planet ACF\label{fig1a}]{%
  \includegraphics[width=.35\linewidth]{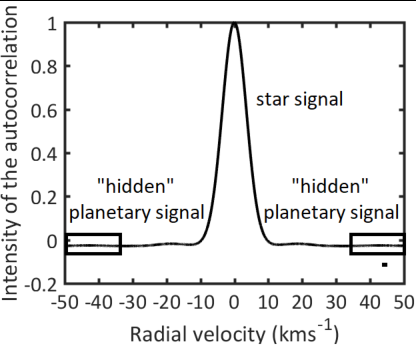}%
}
\subfloat[Star-only ACFs template\label{fig1b}]{%
  \includegraphics[width=.352\linewidth]{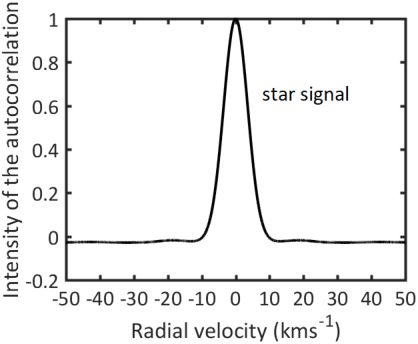}%
  }
\subfloat[Recovered planetary signal\label{fig1c}]{%
  \includegraphics[width=.368\linewidth]{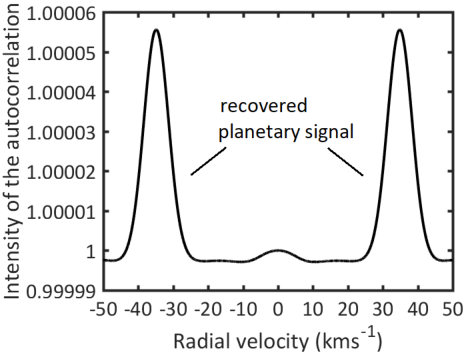}%
  }
\caption{Recovering a planetary signal from a star-only ACFs template. Each star+planet ACF (a) are divided by the star-only ACFs template (b) in order to recover the planetary signal (c).}
\label{fig1}
\end{figure*}

We begin by removing the continuum of each spectrum before applying the ACF to each spectrum. The objective is to isolate the signal of the spectral lines and eliminate any contribution of the continuum in the ACF profile. By doing so, we ensure to measure only the intensity due to the signal of the star and the planet in the ACF profile and not variations of intensity in the continuum of the spectrum.  After removing the continuum and applying the ACF, we observe a prevailing star signal in the ACF profile bounded on each side by a continuum zero intensity (see Fig. \ref{fig1a}). The removal of the continuum is done by using the function smooth of MATLAB. This function applies a smoothing to the input vector (i.e stellar spectrum) using a mobile filter whose size is determined by the user. For example, a mobile filter of 3 pixels would yield the following intensity values as a function of the $i^\textrm{th}$ pixel of the velocity axis: \\

$\textrm{I(1)}_\textrm{smooth} = \textrm{I(1)}$, 

$\textrm{I(2)}_\textrm{smooth} = [\textrm{I(1)} + \textrm{I(2)} + \textrm{I(3)}]/3$, 

$\textrm{I(3)}_\textrm{smooth} = [\textrm{I(2)} + \textrm{I(3)} + \textrm{I(4)}]/3$, and so forth. \\

\noindent In this work, we use a mobile filter that has a width equal to 0.5 \% of the total number of pixels of each spectrum. The smoothed version of the spectrum resulting from the application of the mobile filter thereby represents the continuum. The mobile filter size that we chose allowed us to effectively isolate the continuum of each spectrum without removing any other important information inherent in the spectrum. The last step in removing the continuum consists in normalizing each spectrum by their corresponding smoothed variant. To do so, we subsequently divide each spectrum by their smoothed version (i.e. the continuum) obtained with the application of the mobile filter. 

We apply the ACF separately to each of the 20 spectra that do not contain a planetary signal. Afterwards, we coadd the ACF of these 20 spectra to create a star-only ACFs template for the signal of the star with a higher SNR than the autocorrelation of each spectrum taken individually. The peak intensity of the signal of the star (and that of the planet) in the ACF profile depends on the SNR of the spectrum. It should be noted that dividing an ACF containing a planetary signal by the star-only ACFs template would lead to a calculation error on each side of the star signal in the ACF profile due to the continuum of zero intensity. To address this, we add a constant to each ACF profile whose value is the peak intensity of the star signal. As an example, if ACF \#1 has a peak intensity of 50 and ACF \#2 has a peak intensity of 25, we would add the value of 50 to ACF \#1 and the value of 25 to ACF \#2. ACF \#1 would then have a peak intensity value of 100 and a continuum value of 50 on both side of the signal. ACF \#2 would have a peak intensity value of 50 and a continuum value of 25. This allows us to work around the division by 0 issue and keep the relative peak intensity value between each ACF.

We autocorrelate, one after the other, each of the 10 spectra containing a planetary signal and divide it by the star-only ACFs template created in the previous step. It then becomes possible to remove the signal of the star and recover the very weak signal of the planet. We illustrate this step in \figref{fig1}. We show an ideal case where no photon noise was added to the spectra. Since the ACF is symmetrical, the signal of the planet can be found in both the positive and negative values at a position equidistant from the value 0. Note also that, unlike the CCF, the planetary signal is characterized by a Gaussian form of positive amplitude when measured with the ACF.

The radial velocity values at which the observed planetary signal is located with the ACF are determined by the following equation:
\begin{equation}
RV_\textrm{planet} = \pm K_{\textrm{planet}} sin(2\pi\phi)
\label{eq2.4}
\end{equation}

\noindent where $\textrm{RV}_\textrm{planet}$ is the observed radial velocity of the planetary signal, $K_{\textrm{planet}}$ is the orbital velocity semi-amplitude of the planet and $\phi$ is the orbital phase of the planet.

For example, if the planet is in phase $\phi $ = 0 or 0.5 (which corresponds respectively to the moment when the planet is in transit in front of the star and when the planet is behind the star with respect to the observer), the observed radial velocity of the planetary signal would be of 0 \kms. At these phases, there is an overlap between the signal of the star and the planet. When the planet is at an orbital phase different from $\phi $ = 0 or 0.5, its signal is shifted from that of the star by a radial velocity value determined by Eq. \eqref{eq2.4}. By using Eq. \eqref{eq2.4}, we calculate the radial velocity with respect to the center-of-mass of the system rather than with respect to the star. However, since the velocity of the star is so low compared to that of the planet, this approximation is valid.

The 10 simulated spectra that contain a planetary signal have a different orbital phase. We thus obtain, after dividing each of these spectra by the star-only ACFs template, 10 planetary signals that do not overlap at the same radial velocity values (positive and negative). In order to maximize the signal of the planet, we must superimpose the 10 planetary signals. To do this, we respectively shift the left side (negative velocity values) and the right side of the ACF profile (positive velocity values) to the center value of 0. These shifts are done according to Eq. \eqref {eq2.4}.

The precision of this step depends on the velocity increment per pixel of the interpolation. When the correct orbital velocity semi-amplitude value ($K_{\textrm{planet}}$) of the planet is selected, the 10 planetary signals will overlap at the center value of 0 which will maximize the SNR of the measured planetary signal. If the orbital velocity semi-amplitude value of the planet is overestimated or underestimated, the superposition of the planetary signals will not be adequate and the detection significance will be reduced.

The strength of the resulting planetary signal is defined as the amplitude of the signal at the center value of 0 over the standard deviation of the continuum noise next to the signal of the planet:

\begin{equation}
D = \frac{A}{\sigma_\textrm{noise}},
\label{eq4}
\end{equation}

\noindent where $D$ is the detection significance of the planetary signal, $A$ is the amplitude the planetary signal and $\sigma_\textrm{noise}$ is the standard deviation of the continuum noise next to the signal of the planet. Notations are the same ones used by \cite{martins2015}.  

The CCF is prone to systematic effects and the ACF is no exception. Therefore, the continuum noise that is measured next to the planetary signal is a blend of random noise and systematic effects. In Eq. \eqref{eq4}, the $\sigma_\textrm{noise}$ term is defined as the standard deviation of the pixel intensity next to the detected planetary signal. We use the same definition as \cite{martins2015} for the evaluation of the continuum noise. This allow us to establish a direct comparison between the two techniques.

Because the ACF is symmetrical and relies on a double radial velocity shift technique, it is important to evaluate the noise far enough from the detected planetary signal. Otherwise, the evaluation of the noise could be slightly mistakenly evaluated because of the presence of multiple planetary signals as a result of the radial velocity shifts. The noise in the ACF profile is only symmetrical with respect to the central position. However, the isolated planetary signals obtained after dividing each star + planet ACF by the star-only ACFs template are of a Gaussian form and therefore symmetrical. The intensity of both Gaussian planetary signals on each side of the radial velocity axis will coadd at the central position due to their symmetrical nature while the noise will be attenuated. Consequently, we increase the SNR of the planetary signal at the central position when applying double radial velocity shifts over a single one.

\section{Results}
\label{results}

In this section, we first compare the performance of both the ACF and the CCF in the attempt to recover a planetary signal with a known orbital velocity semi-amplitude of 125 \kms using simulated spectra. Afterwards, we analyse the 51 Peg + 51 Peg b system with the ACF and compare ours results with those obtained by \cite{martins2015} who used the CCF method.

\subsection{Detection of a planetary signal with simulated spectra}
\label{simulated}

The results presented in this section are based on simulations conducted with the ACF and CCF methods. In total, 3 simulations are run with different parameters (shown below) for the 10 ACFs containing a planetary signal with a predetermined orbital velocity semi-amplitude $K_{\textrm{planet}}$ of 125 \kms. We assess and compare the impact of the precision of the interpolation when using the ACF and the CCF.   For each simulation run, we show the results obtained with the ACF and the CCF using a velocity increment per pixel of 50 \ms, 10 \ms and 1 \ms for the interpolation. \\

(i) First run: the 10 ACFs with a planetary signal have an orbital phase $\phi$ between 0.41 and 0.455, no noise is added to the spectra.

(ii) Second run: same as first run but with noise added to all 30 spectra.

(iii) Third run: the 10 ACFs with a planetary signal have an orbital phase $\phi$ between 0.445 and 0.49, no noise is added to the spectra.

The reason why we chose these particular orbital phase intervals in the simulation runs is discussed below.

The noise is simulated with MATLAB software using the function randn. The amount of added noise is set so that the standard deviation of the noise is on average around 40 times greater than the amplitude of the planetary signal. This gives the spectra with simulated random noise a SNR of about 1500. The reason why the SNR is so high for these simulated spectra is because (a) fewer lines are simulated in comparison to real spectra and (b) fewer spectra are used in comparison to real data analysis for the detection of a planetary signal. Mathematically, we would obtain equivalent results if we used more lines in our simulations with a lower SNR for each spectrum. The parameters for our simulated spectra (quantity of lines, quantity of spectra) were chosen in order to decrease the computation time because the simulations take a considerable amount of computation time. The high SNR compensates for the small quantity of simulated spectra and the number of lines used. As mentioned in Section \ref{generate}, both the ACF and the CCF behave similarly to the number of lines involved in the data analysis. Therefore, the number of lines used in the simulations does not benefit or disadvantage one technique more than the other. For both techniques, we used 600 lines with a FWHM of 7.50 \kms ~for the signal of the star. Moreover, we used a weighted binary mask for the CCF technique. For all runs, the planetary signal was recovered using the known $K_{\textrm{planet}}$ value of 125 \kms.

\begin{figure*}
\subfloat[Interpolation velocity increment per pixel $\Delta \textrm{v}$ = 50 \ms. \label{fig4_50}]{%
  \includegraphics[width=1\linewidth]{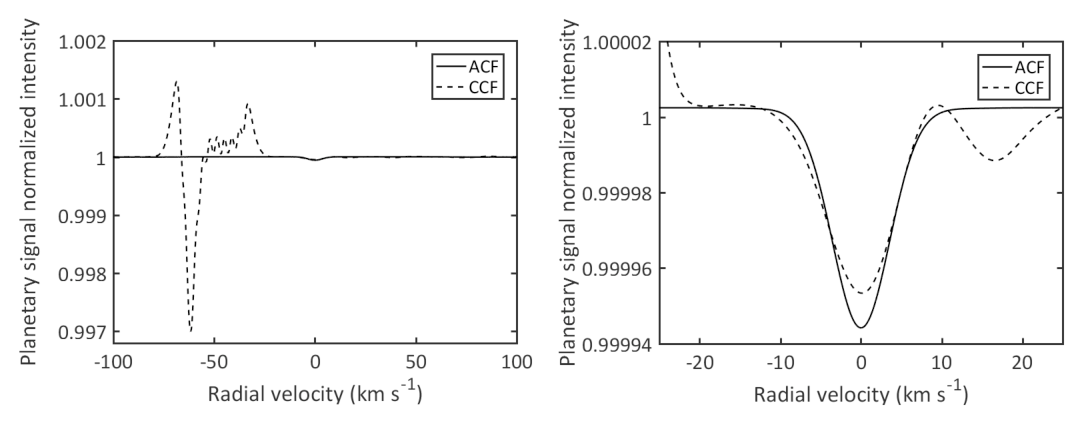}%
}

\subfloat[Interpolation velocity increment per pixel $\Delta \textrm{v}$ = 10 \ms. \label{fig4_10}]{%
  \includegraphics[width=1\linewidth]{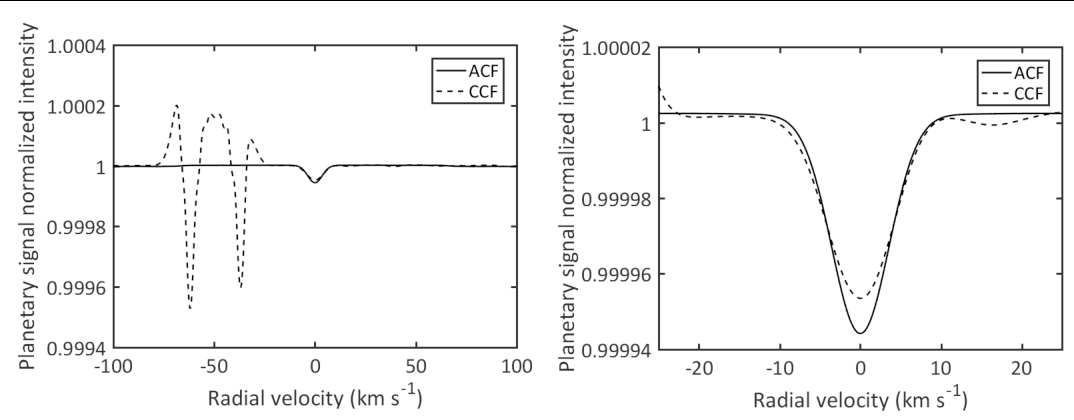}%
}

\subfloat[Interpolation velocity increment per pixel $\Delta \textrm{v}$ = 1 \ms. \label{fig4_1}]{%
  \includegraphics[width=1\linewidth]{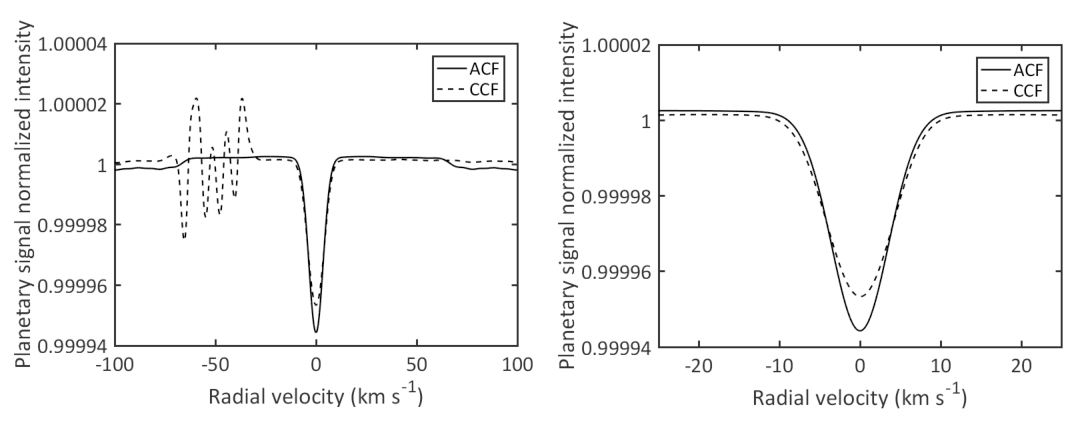}%
}

\caption{First simulation run without random noise. Recovered planetary signals at $K_{\textrm{planet}}$ = 125 \kms ~using simulated spectra. The 10 ACFs/CCFs with a planetary signal have an orbital phase $\phi$ between 0.41 and 0.455. (Left) whole signal observed between -100 and +100 \kms. (Right) zoomed version of the left figure where the recovered planetary signal is observed. }
\label{fig4}
\end{figure*}

\figref{fig4} shows the results obtained for the first simulation run. To facilitate the comparison between the two methods, the planetary signal measured with the ACF was flipped upside down. On the left side of \figref{fig4}, we show the whole signal obtained between -100 and +100 \kms for different velocity increment per pixel values. On the right side, we display a zoomed version of the left figures between -25 and +25 \kms where the recovered planetary signal is observed at the central position. We note that the characteristics of the planetary signal (i.e. the shape, the amplitude and the FWHM) using the ACF is the same regardless of the velocity increment per pixel value. This illustrates the precision and the efficiency of the ACF at lower interpolation resolutions. On the other hand, the results obtained with the CCF improved as we increased the precision of the interpolation. The shape of the planetary signal is affected at $\Delta \textrm{v} = 50$ \ms when measured by the CCF, especially in the wings of the signal. The amplitude of the artifacts originating from the CCF normalization by the template also decreased significantly as we lowered the velocity increment per pixel value. The simulated spectra for the first and the second run have an orbital phase $\phi$ between 0.41 and 0.455. We used this particular phase interval for the first two simulation runs because it is representative of the one used by \cite{martins2015} for the analysis of 51 Peg b with the CCF. 

\begin{figure*}
\subfloat[Interpolation velocity increment per pixel $\Delta \textrm{v}$ = 50 \ms. \label{fig5_50}]{%
  \includegraphics[width=1\linewidth]{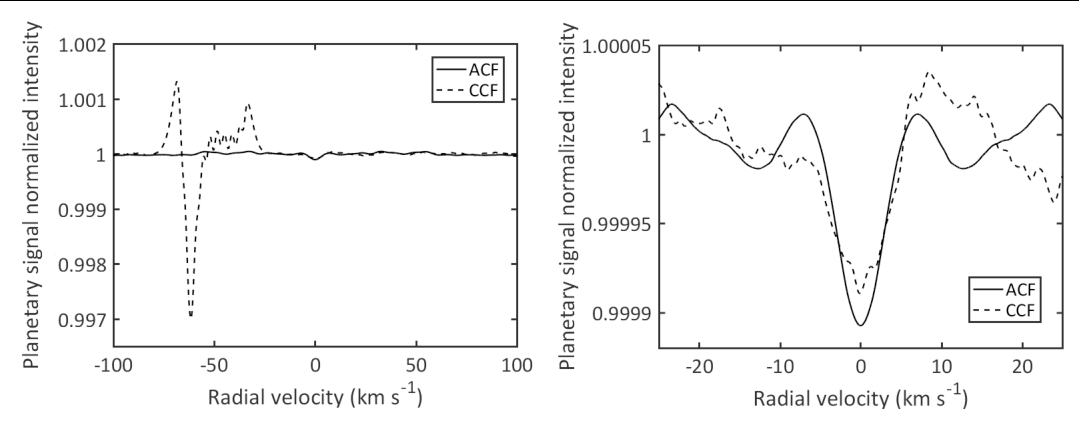}%
}

\subfloat[Interpolation velocity increment per pixel $\Delta \textrm{v}$ = 10 \ms. \label{fig5_10}]{%
  \includegraphics[width=1\linewidth]{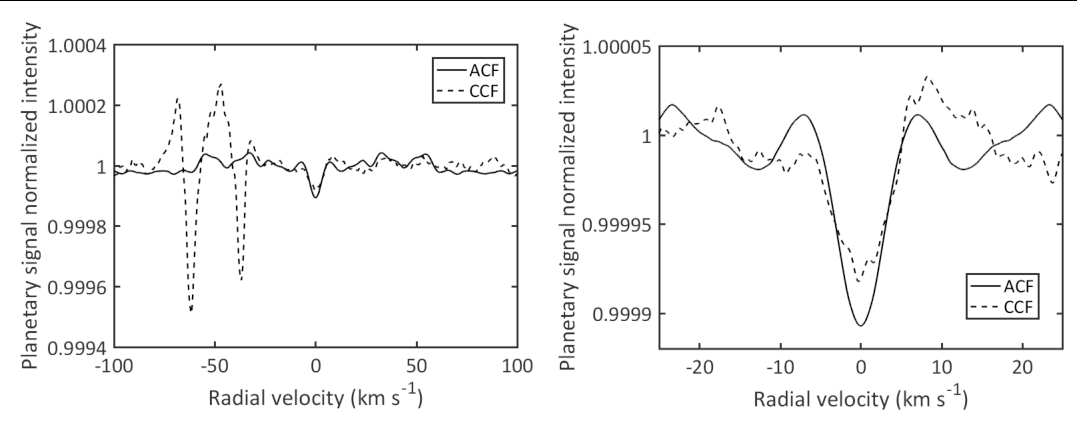}%
}

\subfloat[Interpolation velocity increment per pixel $\Delta \textrm{v}$ = 1 \ms. \label{fig5_1}]{%
  \includegraphics[width=1\linewidth]{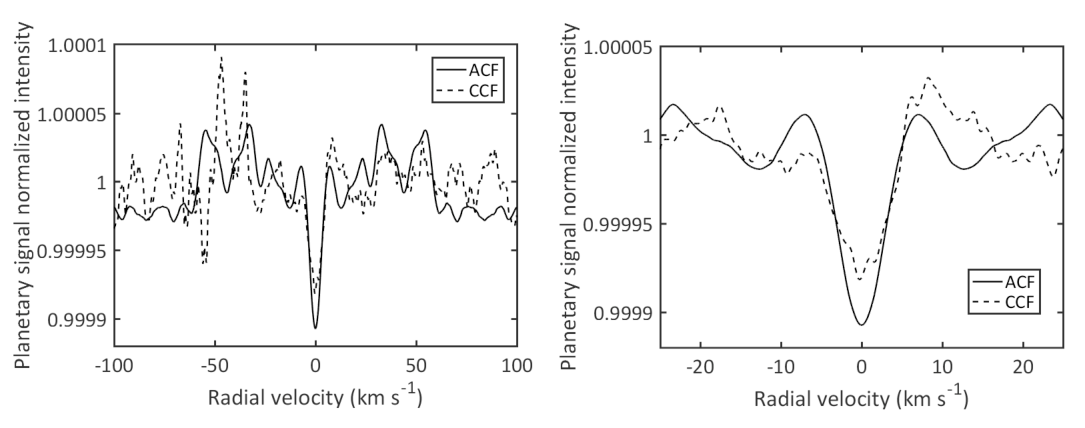}%
}

\caption{Second simulation run with added random noise. Recovered planetary signals at $K_{\textrm{planet}}$ = 125 \kms ~using simulated spectra. The 10 ACFs/CCFs with a planetary signal have an orbital phase $\phi$ between 0.41 and 0.455. (Left) whole signal observed between -100 and +100 \kms. (Right) zoomed version of the left figure where the recovered planetary signal is observed. }
\label{fig5}
\end{figure*}

\figref{fig5} illustrates the second run. In this run, the same spectra as the first run were used but this time simulated random noise was added. The planetary signal can be observed with both methods albeit with more precision with the ACF. The ACF provides us with similar results regarding the characteristics of the observed planetary signal compared to the first simulation run. This demonstrates the efficiency of the ACF against photon noise. The results obtained for the CCF at $\Delta \textrm{v}$ = 10 and 1 \ms are similar. However, the importance of artifacts is once again diminished when using the lowest velocity increment per pixel value. The first two simulation runs demonstrate that the signal of the planet can be detectable with both the ACF and the CCF methods if positioned in every spectrum far enough from the signal of the star.

\begin{figure*}
\subfloat[Interpolation velocity increment per pixel $\Delta \textrm{v}$ = 50 \ms. \label{fig6_50}]{%
  \includegraphics[width=1\linewidth]{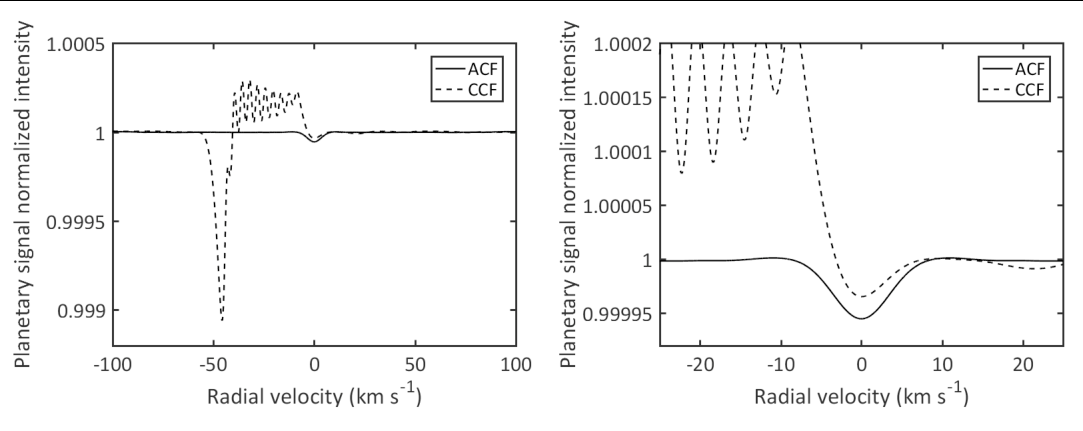}%
}

\subfloat[Interpolation velocity increment per pixel $\Delta \textrm{v}$ = 10 \ms. \label{fig6_10}]{%
  \includegraphics[width=1\linewidth]{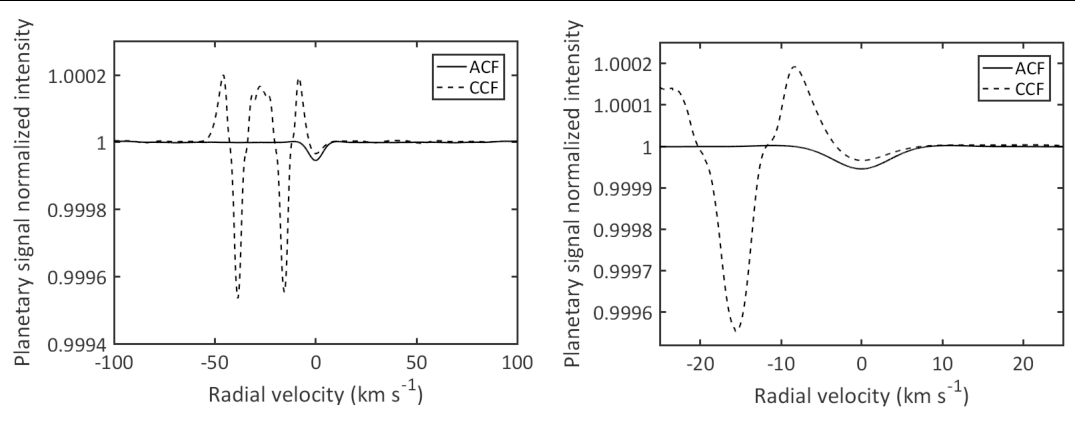}%
}

\subfloat[Interpolation velocity increment per pixel $\Delta \textrm{v}$ = 1 \ms. \label{fig6_1}]{%
  \includegraphics[width=1\linewidth]{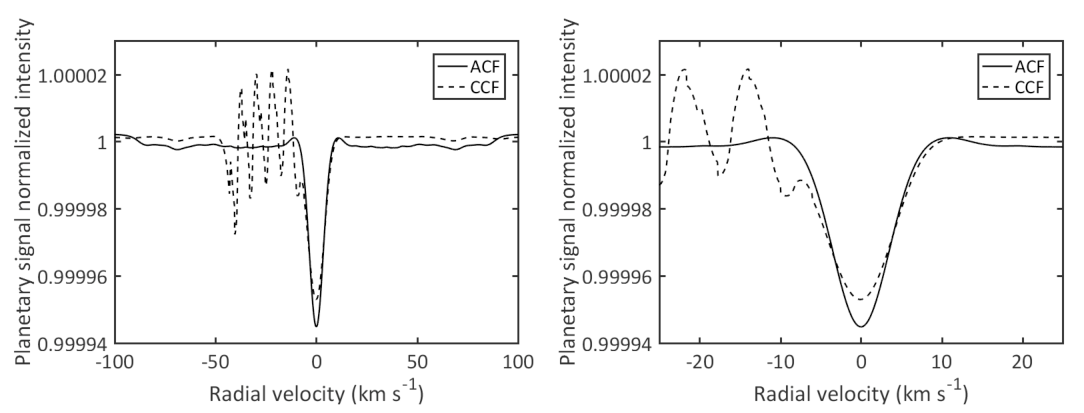}%
}

\caption{Third simulation run without random noise. Recovered planetary signals at $K_{\textrm{planet}}$ = 125 \kms ~using simulated spectra. The 10 ACFs/CCFs with a planetary signal have an orbital phase $\phi$ between 0.445 and 0.49. (Left) whole signal observed between -100 and +100 \kms. (Right) zoomed version of the left figure where the recovered planetary signal is observed. }
\label{fig6}
\end{figure*}

In the third run, the simulated spectra have an orbital phase $\phi$ between 0.445 and 0.49. The reason why we use this phase range for these simulations is to analyse the efficiency of both the ACF and the CCF when using spectra where the planetary signal is close to the signal of the star. Although the ACF provided us with great results in the first two simulations runs, its main advantage comes from the spectra where the planetary signal is closer to the signal of the star. Results for such a case are shown in \figref{fig6}. At these phases, the CCF technique suffers from artifacts polluting the planetary signal. For the first two simulation runs, the planetary signal is sufficiently shifted due to the orbital phase of the spectra. This ensures that the artifacts don't interfere with the signal of the planet when using the CCF. For the third simulation run, the planetary signal is not shifted at a sufficient distance when using the CCF. Therefore, the artifacts completely alter the signal of the planet for a velocity increment value of 50 and 10 \ms. This effect is smaller as the velocity increment value of the interpolation decreases. We note in \figref{fig6_1} that the planetary signal can be recovered at $\Delta \textrm{v}$ = 1 \ms for the CCF. However, this precision of interpolation requires an exceptionally high amount of computation time. 

The planetary signal obtained with the ACF using spectra with an orbital phase $\phi$ between 0.445 and 0.49 can be measured without difficulty for any interpolation precision shown in \figref{fig6}. Therefore, these simulations prove that ACF allows us to use spectra where the signal of the planet and that of the star are very close. This provides a significant advantage for the ACF as more spectra can be used compared to the CCF to recover the planetary signal.

\subsection{The case of 51 Peg b}
\label{51peg}

In this section, we analyse the 51 Peg + 51 Peg b system using the ACF and compare the results obtained with those of \cite{martins2015} who used the CCF. We retrieved data from the ESO database and analysed the same 90 spectra acquired with the HARPS spectrograph that were used by \cite{martins2015}. For this system, we set the velocity increment to 50 \ms per pixel which is the same resolution used by \cite{martins2015}. The orbital parameters used for 51 Peg b are the same as those found in Table 3 of \cite{martins2015}.  

Out of the 90 spectra, 20 are cases where the planet is located near the inferior conjunction (i.e no star light is reflected off of the planet). The position of the planetary signal in the rest of the spectra varies rapidly from spectrum to spectrum due to the small orbital period of the planet (4.231 d). It is therefore advantageous to use all 90 spectra to build the star-only ACFs template as planetary signal withdrawal by the template will be minimal. With the ACF, we obtain a FWHM of $10.60 \pm 0.26$ \kms for the signal of the star on the 90 ACFs template. Dividing the FWHM of the star by the $\sqrt{2}$ multiplicative factor brought mathematically by the ACF on Gaussian functions (see Section \ref{autocorr}), we obtain a $\textrm{FWHM}_{\textrm{star}}$ = $7.50 \pm 0.18$ \kms.

As mentioned in Section \ref{detecting}, measuring noise further away from the central position where the planetary signal is detected ensures that the evaluated noise does not contain planetary signals when using the ACF. For 51 Peg b, we measured the standard deviation of the pixel intensity between 200 and 700 \kms. Covering a large interval (500 \kms in this case) allows us to obtain a precise and representative measurement of the continuum noise. The intensity of the continuum noise in the recovered planetary signal (see Fig. \ref{fig1c}), when using the ACF, is maximal at radial velocity v = 0 \kms and slowly decreases gradually on each side of (positive and negative values) of the central position at very high values. However, in this work, the continuum noise is evaluated close enough to the central position for this intensity decrease to be completely negligible. This allows us to make sure that the continuum noise next to the planetary signal is not underestimated and properly evaluated.

\begin{figure*}
\subfloat[Detected planetary signal for the $8 \times \textrm{FWHM}_{\textrm{star}}$ condition at 133 \kms. (Left) Detected planetary signal between -125 and +125 \kms. (Right) Evaluated noise between 200 and 700 \kms.  \label{fig6a}]{%
  \includegraphics[width=0.95\linewidth]{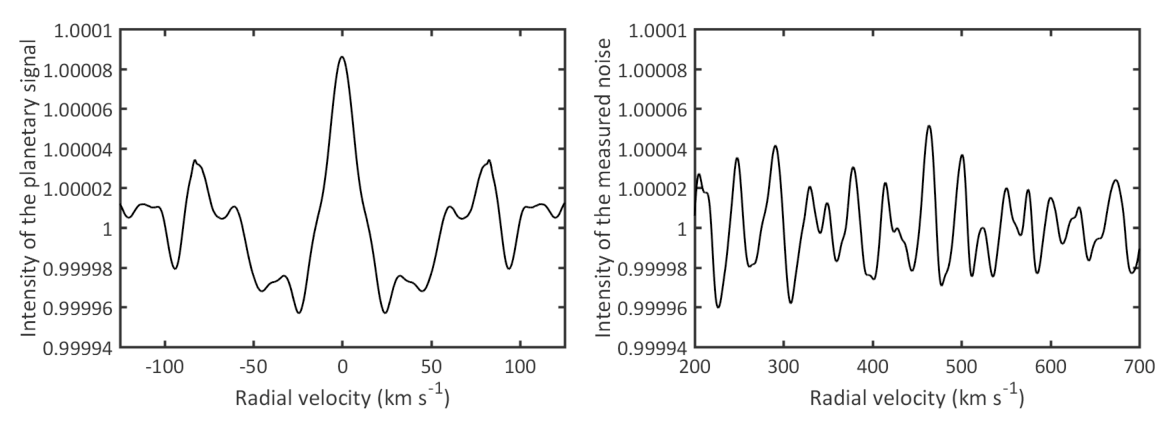}%
}

\subfloat[Detected planetary signal for the $7 \times \textrm{FWHM}_{\textrm{star}}$ condition at 132 \kms. (Left) Detected planetary signal between -125 and +125 \kms. (Right) Evaluated noise between 200 and 700 \kms. \label{fig6b}]{%
  \includegraphics[width=0.95\linewidth]{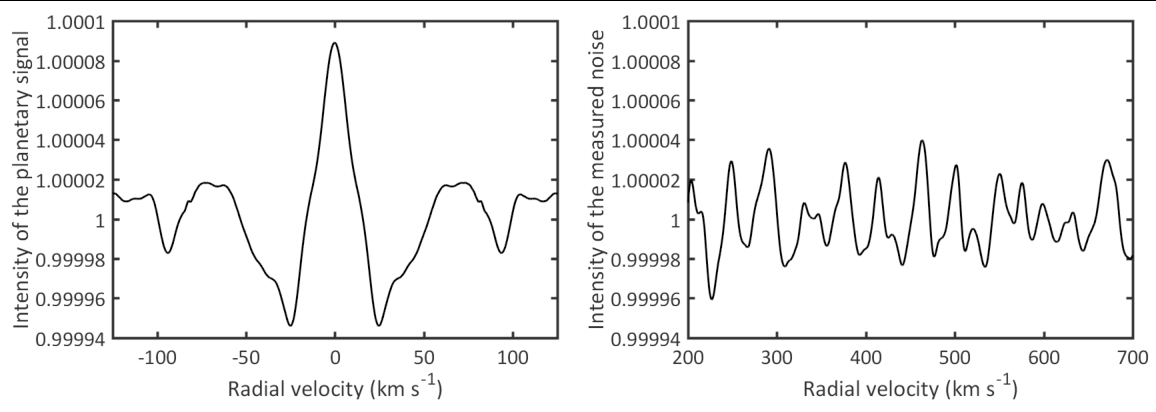}%
}

\caption{Recovered planetary signal for the $8 \times \textrm{FWHM}_{\textrm{star}}$ and $7 \times \textrm{FWHM}_{\textrm{star}}$ conditions. }
\label{fig7}
\end{figure*}

\begin{figure*}

  \includegraphics[width=0.68\linewidth]{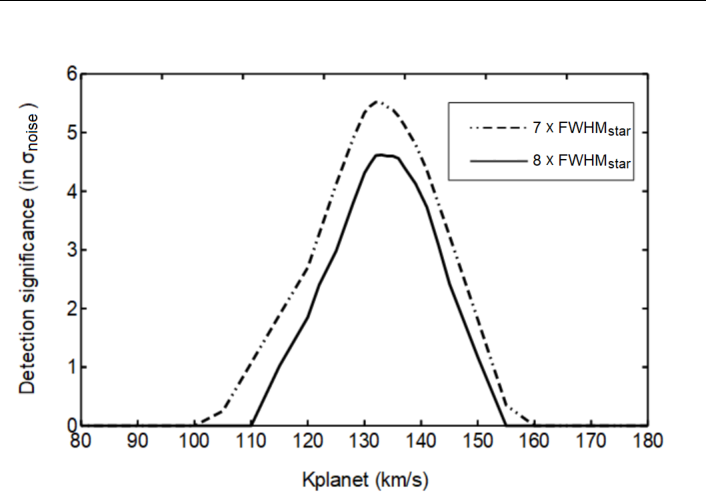}%

\caption{Detection significance of the recovered planetary signal when using 25 spectra ($8 \times \textrm{FWHM}_{\textrm{star}}$) and 36 spectra ($7 \times \textrm{FWHM}_{\textrm{star}}$).}
\label{fig8}
\end{figure*}

\begin{figure*}

  \includegraphics[width=1\linewidth]{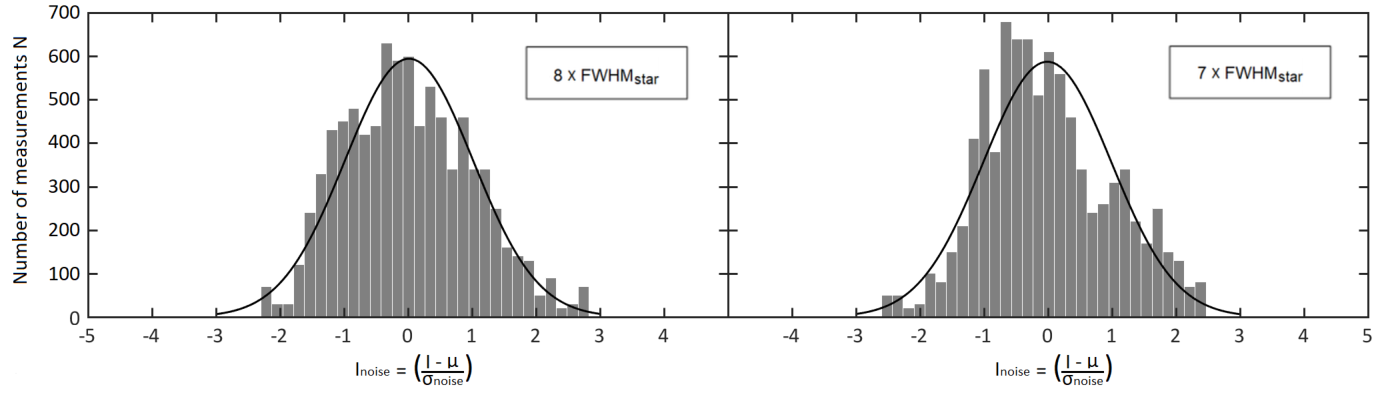}%

\caption{Distribution of $I_{\textrm{noise}}$ between 200 and 700 \kms for the $8 \times \textrm{FWHM}_{\textrm{star}}$ condition at 133 \kms (left) and for the $7 \times \textrm{FWHM}_{\textrm{star}}$ condition at 132 \kms (right)}
\label{fig8hist}
\end{figure*}

We first conducted a direct comparison between the two techniques by using only spectra that contain a planetary signal with a velocity gap of at least $8 \times \textrm{FWHM}_{\textrm{star}}$ relatively to the signal of the star. This is the gap that was used by \cite{martins2015} with the CCF. The parameters of 51 Peg b are well documented and its orbital velocity semi-amplitude $K_{\textrm{planet}}$ of about 135 \kms                                                                                                                                                                                                                                                                                                                                                                                                                                                                                                                                           is known \citep{brogi2013,martins2015,birkby2017}. Our search for the $K_{\textrm{planet}}$ value is confined around the known orbital velocity of the planet. We therefore attempt to recover the planetary signal for different values of orbital velocity semi-amplitude $K_{\textrm{planet}}$ between 80 and 180 \kms with a 1 \kms step increment. A maximal detection significance of 4.62 $\sigma_\textrm{noise}$ was achieved at a $K_{\textrm{planet}}$ of 133 \kms. Uncertainty values associated to all parameters of the planetary signal were set by calculating the standard deviation between the signal and a least-square Gaussian fit. We measure a FWHM of $13.70 \pm 0.39$ \kms and an amplitude of $8.62 \pm 0.20 \times 10^{-5}$ for the detected signal. This yields a $\textrm{FWHM}_{\textrm{planet}}$ of $9.69 \pm 0.28$ \kms, after correcting for the $\sqrt{2}$ multiplicative factor, which is roughly 30\% higher than the $\textrm{FWHM}_{\textrm{star}}$ value we measured. Compared to the CCF technique used by \cite{martins2015}, we made a detection with a signal-to-noise ratio that is about 25\% higher with the ACF at a very similar orbital velocity semi-amplitude value using the same spectra.

Another important point to mention relates to the physical appearance of the signal measured with the ACF. The signal of the planet that we obtained has a clear and distinctive Gaussian form (see Fig. \ref{fig6a}) in comparison to the one in Figure 5 of \cite{martins2015} using the CCF. Reasons for this include the double radial velocity shift technique of the ACF which gives us a symmetrical planetary signal. Also, the Gaussian nature of the lines in the mask of the ACF, which is the spectrum itself, strengthens the Gaussian appearance of the obtained planetary signal. Uncertainty values are therefore lower when using the ACF considering the near perfect match between the Gaussian fit and the planetary signal.

The signal of the planet that we obtained has an amplitude whose order of magnitude agrees well with what was obtained by \cite{martins2015} using the CCF. However, the $\textrm{FWHM}_{\textrm{planet}}$ value of $9.69 \pm 0.28$ \kms that we measured is considerably lower than the value of $22.6 \pm 3.6$ \kms obtained by \cite{martins2015}. They speculated that a rapid rotational motion of the planet could potentially be the cause of this strong broadening in the planetary signal. Contrary to \cite{martins2015}, we do not identify rapid differential rotation as a potential dominant broadening mechanism regarding the FWHM of the planetary signal. The small difference between our $\textrm{FWHM}_{\textrm{star}}$ ($7.50 \pm 0.18$ \kms) and $\textrm{FWHM}_{\textrm{planet}}$ ($9.69 \pm 0.28$ \kms) values in our measurements is presumably attributed mainly to the noise polluting the weak planetary signal.

One of the important advantages of the ACF in analysing planetary systems is the ability to use spectra at phases where the signal of the planet is closer to the signal of the star.  Therefore, we decided to lower the condition for the use of spectra to a still somewhat conservative value of $7 \times \textrm{FWHM}_{\textrm{star}}$. In doing so, we detected the signal of the planet with a maximal detection significance of 5.52 $\sigma_\textrm{noise}$ at a $K_{\textrm{planet}}$ of 132 \kms (see Fig. \ref{fig6b}). The measured $\textrm{FWHM}_{\textrm{planet}}$ is $10.40 \pm 0.25$ \kms after correcting for the $\sqrt{2}$ multiplicative factor. The set of parameters of the detected planetary signal for both conditions can be found in Table \ref{table1}.

In all, 25 spectra were used for the $K_{\textrm{planet}}$ of maximal detection significance for the $8 \times \textrm{FWHM}_{\textrm{star}}$ condition while 36 spectra were used for the $7 \times \textrm{FWHM}_{\textrm{star}}$ one. \figref{fig8} shows the detection significance as a function of $K_{\textrm{planet}}$ for both conditions. Assuming spectra of similar SNR on average, we should expect an improvement of about 20\% ($\sqrt{36/25}$) in the detection significance of the planetary signal when using 36 spectra over 25. Our results are in agreement with this hypothesis. 

One may of course wonder whether the pixel intensity next to the planetary signal that we obtained with the ACF follows a Gaussian distribution. We carried out work to verify it. This is discussed in the next sentences.

\begin{figure*}

  \includegraphics[width=1\linewidth]{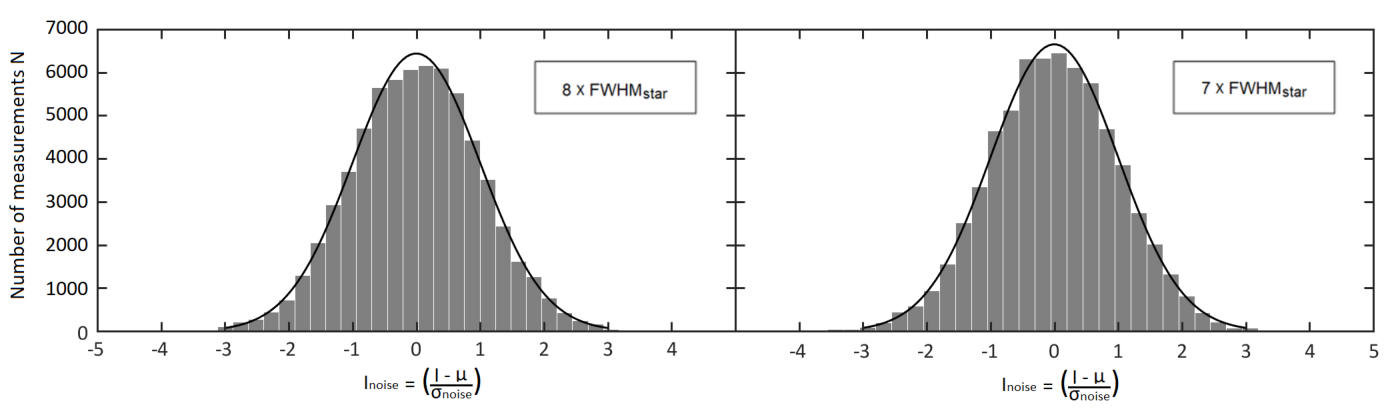}%

\caption{Distribution of $I_{\textrm{noise}}$ between 200 and 5000 \kms for the $8 \times \textrm{FWHM}_{\textrm{star}}$ condition at 133 \kms (left) and for the $7 \times \textrm{FWHM}_{\textrm{star}}$ condition at 132 \kms (right)}
\label{fig10hist}
\end{figure*}

For each pixel between 200 and 700 \kms, we calculated the difference from the mean normalized by the standard deviation $\sigma_\textrm{noise}$. This is defined as $I_{\textrm{noise}_i} = \frac{I_i - \mu}{\sigma_\textrm{noise}}$, where $I_i$ is the intensity of the $i^{th}$ pixel and $\mu$ is the mean pixel intensity in the interval considered. We subsequently made a histogram of the number of measurements N as a function of $I_{\textrm{noise}}$. Figure \ref{fig8hist} shows the distribution of $I_{\textrm{noise}}$ between 200 and 700 \kms for the $8 \times \textrm{FWHM}_{\textrm{star}}$ condition at 133 \kms (left) and for the $7 \times \textrm{FWHM}_{\textrm{star}}$ condition at 132 \kms (right). The best Gaussian fit was added to each distribution. Since both histograms (i.e. $8 \times \textrm{FWHM}_{\textrm{star}}$ and $7 \times \textrm{FWHM}_{\textrm{star}}$ conditions) exhibit an obvious Gaussian distribution, it becomes possible to calculate the probability that a detection of 4.62 $\sigma_\textrm{noise}$ is part of the distribution of the pixel intensity next to the planetary signal (i.e. the probability of having a false positive planetary detection). Under the assumption that the pixel intensity next to the measured planetary signal does follow a Gaussian probability density function, there is a 1 in 260,000 chance of having a false detection with the ACF technique (for a 4.62 $\sigma_\textrm{noise}$ detection) while, with the CCF, there is a 1 in 4,600 chance of having a false detection (for a 3.70 $\sigma_\textrm{noise}$ detection). This is true when using the same number of spectra as \cite{martins2015} respecting the $8 \times \textrm{FWHM}_{\textrm{star}}$ condition. Figure \ref{fig8hist} demonstrates that the pixel intensity between 200 and 700 \kms follows a Gaussian distribution and validates our statistical analysis. Note that on first glance, the distributions on Figure \ref{fig8hist} appear partly skewed, especially the one on the right-hand side referring to the $7 \times \textrm{FWHM}_{\textrm{star}}$ condition. This is mostly due to the limited interval used (i.e. between 200 and 700 \kms) for the evaluation of $\sigma_\textrm{noise}$ and hence, the lack of data to create statistically significant histograms. Small residuals originating from the removal of the continuum could also create asymmetries in the distributions. Figure \ref{fig10hist} shows the distributions of the pixel intensity that we obtain when using an interval between 200 and 5000 \kms. Using a larger interval allows us to recover fully symmetrical Gaussian histograms for both the $8 \times \textrm{FWHM}_{\textrm{star}}$ and the $7 \times \textrm{FWHM}_{\textrm{star}}$ conditions.

\begin{figure*}

  \includegraphics[width=0.68\linewidth]{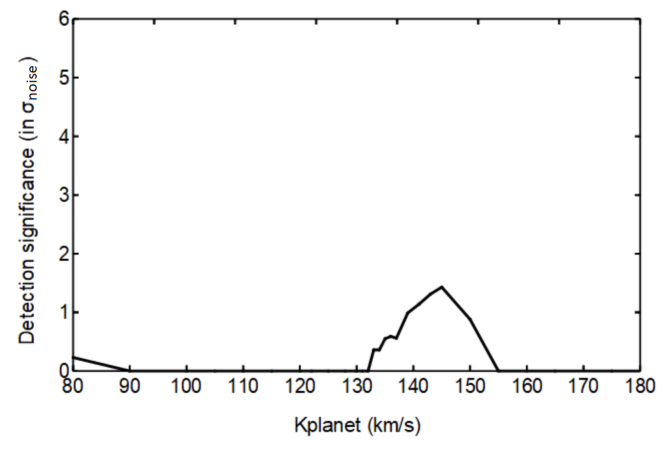}%

\caption{Detection significance of the signal obtained when using 20 spectra where no planetary signal is expected.}
\label{fig9}
\end{figure*}

We perform an additional analysis in what follows to ensure that the technique used to measure the planetary signal is valid. We evaluate the possibility that the detection is the result of a sum of systematic effects that were shifted at the central position of the velocity axis where the planetary signal is analysed. We also consider the possibility of having a false detection due to artifacts generated by the removal of the continuum. Theoretically, no significant detection should be achieved when using spectra at phases where the planet is at inferior conjunction (i.e. no stellar light is reflected by the planet with respect to Earth). To assess the likelihood of having systematic effects or continuum artifacts as a false detection, we use the 20 spectra of 51 Peg obtained near phase $\phi$ = 0 and try to recover a planetary signal between $80 \leq K_{\textrm{planet}} \leq 180$ with phases varying between $0.41 \leq \phi \leq 0.455$. Results related to the detection significance of this analysis are shown in \figref{fig9}. We note that, as expected, no significant detection is made when using the 20 spectra at phase $\phi \simeq$  0. We obtain a maximum detection significance of 1.43 $\sigma_\textrm{noise}$ at a $K_{\textrm{planet}}$ of 145 \kms. The significance of this detection is well below the 3 $\sigma_\textrm{noise}$ threshold and therefore negligible. These results imply that the possibilities of having false positive detections due to systematic effects or the removal of the continuum are unlikely.

\begin{figure*}

  \includegraphics[width=1\linewidth]{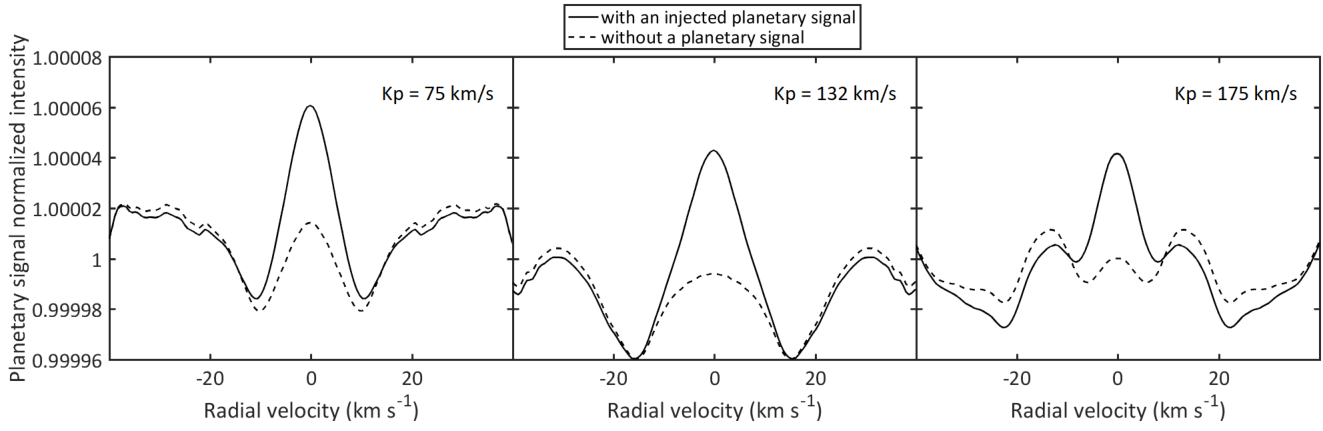}%

\caption{Attempt to recover a planetary signal at 75, 132 and 175 \kms where a planetary signal was injected and where one was not. Full line represent the case where a planetary signal was injected at 75 \kms (left figure), 132 \kms (middle figure) and 175 \kms (right figure). Dotted line is the result obtained when no planetary signal is added.}
\label{fig10}
\end{figure*}

Furthermore, we proceed with more simulations to assess the capabilities of the ACF in detecting a planetary signal by means of reflected light. To this end, we used the 20 spectra of 51 Peg obtained near phase $\phi$ = 0 and ran 3 simulations in which we injected a planetary signal at a known $K_{\textrm{planet}}$ values. The values for the injected planetary signal are $K_{\textrm{planet}}$ = 75 \kms for the first simulation, 132 \kms for the second one and 175 \kms for the third one. The simulated planetary signal that we injected in the spectra are a copy of the spectra of the star. The intensity of the lines were reduced to obtain a planetary signal that is $5 \times 10^{-5}$ that of star. The lines of the planetary signal were then shifted with respect to those of the star so that the phase for the 20 spectra varies between $0.41 \leq \phi \leq 0.455$. We attempt to recover the injected planetary signal at their known $K_{\textrm{planet}}$ values for the 3 simulations. We show the results obtained in \figref{fig10}. For each simulation, we compare the results obtained between the case where a planetary signal was injected and the case where no planetary signal was injected. In all cases, the injected planetary signal was successfully recovered at its real $K_{\textrm{planet}}$ value (3.08 $\sigma_\textrm{noise}$ for the first simulation at 75 \kms, 3.59 $\sigma_\textrm{noise}$ for the second simulation at 132 \kms and 3.79 $\sigma_\textrm{noise}$ for the third simulation at 175 \kms). On the other hand, no significant detection was achieved at these $K_{\textrm{planet}}$ values for each simulation when no planetary signal was added. These results are consistent with our expectations and corroborate the validity of the technique used.

\begin{table}
\begin{tabular}{|l|c|c|}
  \hline \hline
Parameters & Planetary signal \#1 & Planetary signal \#2\\ 

  \hline \hline
  Condition & $8 \times \textrm{FWHM}_{\textrm{star}}$ & $7 \times \textrm{FWHM}_{\textrm{star}}$ \\
  $K_{\textrm{planet}}$ [\kms] & 133  & 132 \\
  Amplitude [-] & $8.62 \pm 0.20 \times 10^{-5}$ & $8.90 \pm 0.10 \times 10^{-5}$ \\
  Detection [$\sigma_\textrm{noise}$] & 4.62 & 5.52\\
  \textrm{FWHM} [\kms] & $9.69 \pm 0.28$  &  $10.40 \pm 0.25$\\
  \hline \hline

\end{tabular}
\caption{Set of parameters for the measured planetary signal for both the $8 \times \textrm{FWHM}_{\textrm{star}}$ and $7 \times
\textrm{FWHM}_{\textrm{star}}$ conditions.}
\label{table1}
\end{table}

\section{Discussion and conclusion}
\label{conclusion}

In this paper, we discuss a new method to find and study planets that applies the autocorrelation function (ACF) to spectra of star + planet systems in the optical band. The planetary spectrum is generated by its reflection of the spectrum of the star. We discuss in Section \ref{whyuseacf} the different benefits of using the ACF and how it can overcome some restrictions that the cross-correlation function (CCF) faces when analysing planetary systems. The ACF is a simpler and less time consuming method to use compared to the CCF because no weighted binary mask is required to use the ACF. Also, the precision of the cross-correlation technique depends directly on the velocity increment per pixel of the interpolation. If the velocity increment is not low enough, it is possible that there will be a slight radial velocity offset of less than one pixel after correcting each CCF by the radial velocity of the system. This will lead to the presence of additional artifact near the signal of the star when dividing the CCFs containing a planetary signal by the star-only CCFs template.

The ACF is not affected by this particular problem because the mask is the spectrum itself and therefore the position of the lines in the mask is the same as the position of each individual line in the spectrum. Consequently, each ACF will perfectly coadd at the central position when building the star-only ACFs template (see Fig. \ref{fig1b}).  

\begin{itemize}[label={–}]
\item This advantage makes the ACF technique a more reliable tool. It also allows us to use spectra in our data processing at phases where the signal of the planet is closer to that of the star.   

\item This provides us with the possibility to include more spectra which contain the signal of the planet and therefore improve the signal-to-noise ratio of the detected planetary signal. This was done for 51 Peg b where we lowered the boundary condition to $7 \times \textrm{FWHM}_{\textrm{star}}$ down from the $8 \times \textrm{FWHM}_{\textrm{star}}$ condition established by \cite{martins2015} for the use of spectra with a planetary signal.

\item This change of condition greatly improved the detection significance of our results and confirmed the efficiency of the ACF. We measured a planetary signal with a signal-to-noise ratio 25 \% higher than what was obtained with the CCF with the same spectra. By using more spectra, we obtained a maximal detection significance of 5.52 $\sigma_\textrm{noise}$ at a $K_{\textrm{planet}}$ of 132 \kms.

\item This detection significance value is an improvement of about 49 \% compared to the CCF. At this detection level, the probability of having a false detection is less than 1 in 29,000,000 compared to the CCF technique where there is a 1 in 4,600 chance of having a false detection. We are therefore confident that the detection of the planetary signal is real.
\end{itemize}

Using values above the 5 $\sigma_\textrm{noise}$ detection significance threshold for the assessment of uncertainty gives us a $K_{\textrm{planet}}$ of $132_{-4}^{+7}$ \kms for the $7 \times \textrm{FWHM}_{\textrm{star}}$ condition. This is in line with the $K_{\textrm{planet}}$ of $132_{-15}^{+19}$ \kms  that was measured by \cite{martins2015} and the $K_{\textrm{planet}}$ of $133_{-3.5}^{+4.3}$ \kms obtained by \cite{birkby2017}. Because of the many similarities between the two techniques, the ACF allows to measure the same physical parameters as those obtained with the CCF when detecting planetary signals by means of a direct detection of reflected light. However, we limit the scope of this paper to the efficiency of the ACF to detect planetary signals. Therefore, we do not determine the physical parameters of the planet (e.g. its mass, its orbital inclination) that result from our detection. Considering the similarity of the orbital velocity semi-amplitude value that we measured compared to previous works, the physical parameters resulting from the analysis of the ACF would be very close to what was previously determined for 51 Peg b with the CCF but with smaller uncertainties.

As discussed in Section \ref{51peg}, additional simulations were made in order to demonstrate that the technique used was valid. We were able to successfully recover the artificial planetary signal that we injected with a detection above 3 $\sigma_\textrm{noise}$ for three different $K_{\textrm{planet}}$ values. On the other hand, no significant detection was achieved between $80 \leq K_{\textrm{planet}} \leq 180$ when using spectra where no planetary signal is present. This analysis allowed us to assert that false positive detections due to systematic effects or continuum residuals are very unlikely.

Our results demonstrate that the ACF is a worthy alternative to the CCF for the detection of exoplanets. It is an easy-to-implement tool with a rapid execution speed that provided a considerable improvement in the detection significance of 51 Peg b in comparison with the CCF method.

\section*{Acknowledgements}

This research was supported by the Natural Sciences and Engineering Research Council of Canada.





\bibliographystyle{mnras}

\bibliography{references} 








\bsp	
\label{lastpage}
\end{document}